\documentclass[12pt]{article}

\usepackage[utf8]{inputenc}
\usepackage{textcomp}
\usepackage{chetnew}
\usepackage{tikz}
\usepackage{amssymb,amsfonts,mathrsfs,dsfont,yfonts,bbm}\usetikzlibrary{calc}
\usepackage{xcolor}
\usepackage{braket}

\newcommand{\CH}{\mathcal{H}}

\newcommand{\CC}{\mathcal{C}}
\newcommand{\CO}{\mathcal{O}}

\newcommand{\CI}{\mathcal{I}}
\newcommand{\CN}{\mathcal{N}}

\newcommand*{\boxcoloro}{orange}
\makeatletter
\newcommand{\boxedo}[1]{\textcolor{\boxcoloro}{%
\tikz[baseline={([yshift=-1ex]current bounding box.center)}] \node [rectangle, minimum width=1ex,rounded corners,draw] {\normalcolor\m@th$\displaystyle#1$};}}
 \makeatother
\newcommand*{\boxcolorr}{red}
\makeatletter
\newcommand{\boxedr}[1]{\textcolor{\boxcolorr}{%
\tikz[baseline={([yshift=-1ex]current bounding box.center)}] \node [rectangle, minimum width=1ex,rounded corners,draw] {\normalcolor\m@th$\displaystyle#1$};}}
 \makeatother
\newcommand*{\boxcolorb}{blue}
\makeatletter
\newcommand{\boxedb}[1]{\textcolor{\boxcolorb}{%
\tikz[baseline={([yshift=-1ex]current bounding box.center)}] \node [rectangle, minimum width=1ex,rounded corners,draw] {\normalcolor\m@th$\displaystyle#1$};}}
 \makeatother
\newcommand*{\boxcolorg}{green}
\makeatletter
\newcommand{\boxedg}[1]{\textcolor{\boxcolorg}{%
\tikz[baseline={([yshift=-1ex]current bounding box.center)}] \node [rectangle, minimum width=1ex,rounded corners,draw] {\normalcolor\m@th$\displaystyle#1$};}}
 \makeatother
 \newcommand*{\boxcolorp}{purple}
\makeatletter
\newcommand{\boxedp}[1]{\textcolor{\boxcolorp}{%
\tikz[baseline={([yshift=-1ex]current bounding box.center)}] \node [rectangle, minimum width=1ex,rounded corners,draw] {\normalcolor\m@th$\displaystyle#1$};}}
 \makeatother
  \newcommand*{\boxcolorc}{cyan}
\makeatletter
\newcommand{\boxedc}[1]{\textcolor{\boxcolorc}{%
\tikz[baseline={([yshift=-1ex]current bounding box.center)}] \node [rectangle, minimum width=1ex,rounded corners,draw] {\normalcolor\m@th$\displaystyle#1$};}}
 \makeatother
  \newcommand*{\boxcolory}{yellow}
\makeatletter
\newcommand{\boxedy}[1]{\textcolor{\boxcolory}{%
\tikz[baseline={([yshift=-1ex]current bounding box.center)}] \node [rectangle, minimum width=1ex,rounded corners,draw] {\normalcolor\m@th$\displaystyle#1$};}}
 \makeatother

\usepackage{amsmath}	% required for `\align' (yatex added)
\begin{document}
\preprint{QMUL-PH-19-02}

\title{Rationalizing CFTs and Anyonic Imprints on Higgs Branches}

\author{Matthew Buican$^{\diamondsuit}$ and Zoltan Laczko$^{\clubsuit}$}

\affiliation{\smallskip CRST and School of Physics and Astronomy\\
Queen Mary University of London, London E1 4NS, UK\emails{$^{\diamondsuit}$m.buican@qmul.ac.uk, $^{\clubsuit}$ z.b.laczko@qmul.ac.uk}}

\abstract{We continue our program of mapping data of 4D $\CN=2$ superconformal field theories (SCFTs) onto observables of 2D chiral rational conformal field theories (RCFTs) by revisiting an infinite set of strongly coupled Argyres-Douglas (AD) SCFTs and their associated logarithmic 2D chiral algebras. First, we turn on discrete flavor fugacities (for continuous flavor symmetries) in a known correspondence between certain unrefined characters of these logarithmic theories and unrefined characters of a set of unitary 2D chiral RCFTs. Motivated by this discussion, we then study 4D Higgs branch renormalization group flows (i.e., flows activated by vevs for which only $su(2)_R\subset su(2)_R\times u(1)_R$ is spontaneously broken) emanating from our AD theories from the point of view of the unitary 2D theories and find some surprises. In particular, we argue that certain universal pieces of the topological data underlying the 2D chiral algebra representations associated with the 4D infrared (IR) theory can be computed, via Galois conjugation, in the topological quantum field theory (TQFT) underlying the unitary ultraviolet (UV) chiral RCFT. The mapping of this topological data from UV to IR agrees with the fact that, in our theories, the moduli spaces we study consist of free hypermultiplets at generic points if and only if the UV TQFT is a theory of abelian anyons.}

\date{January 2019}

\setcounter{tocdepth}{2}
\maketitle
\toc

\newsec{Introduction}
Quantum field theory (QFT) in lower dimensions generally seems richer and less rigid than QFT in higher dimensions. For example, in ${\rm D}<7$ dimensions we readily find many examples of interacting conformal field theories (CFTs), while the situation looks somewhat murkier in ${\rm D}\ge7$ (however, see \cite{Cordova:2018eba}). 
As another example, 2D CFTs readily admit non-supersymmetric exactly marginal deformations, while the situation in ${\rm D}>2$ seems far more constrained. In some sense, the relative richness of lower dimensions is to be expected: we can compactify higher dimensional QFTs, and the geometry and topology of the compactifications then enrich the resulting lower-dimensional theories.

Given this picture, we may expect that when a direct algebraic link exists (without going through a compactification) between certain QFTs in higher dimensions and a subset of QFTs in lower dimensions, this subset of lower dimensional QFTs will be \lq\lq small" in comparison with the full space of lower dimensional theories.

One concrete playground in which to test this idea is given in \cite{Beem:2013sza}\footnote{Similar ideas can also be pursued using the more restricted theories in \cite{Beem:2014kka}.}: classes of protected local operators in 4D $\CN=2$ SCFTs called \lq\lq Schur" operators are related to sets of meromorphic currents generating non-unitary 2D chiral algebras.\footnote{Chiral algebras are the set of symmetries of the, say, left-movers (or right-movers) of 2D CFTs.} While the resulting space of 2D chiral algebras is quite large (e.g., see \cite{Beem:2013sza,Lemos:2014lua,Buican:2015ina,Cordova:2015nma,Buican:2016arp,Nishinaka:2016hbw,Buican:2017fiq,Song:2017oew,Xie:2016evu,Creutzig:2017qyf,Choi:2017nur,Creutzig:2018lbc,Buican:2018ddk,Bonetti:2018fqz,Arakawa:2018egx})---reflecting the diversity of 4D $\CN=2$ SCFTs\footnote{It is not clear that the chiral algebra and its representations uniquely specify the 4D theory, so there may be some coarse-graining involved in this correspondence. Note that even in 2D CFT itself, the left and right chiral algebras and their representations are not always sufficient to specify a 2D CFT (e.g., we can have different permutation modular invariants).}---it is a highly constrained subspace within the space of 2D chiral algebras (e.g., see \cite{Liendo:2015ofa,Buican:2016arp,Beem:2018duj} for some constraints). More simply, if we start from unitary 4D theories, then the corresponding 2D theories should be non-unitary chiral algebras with a hidden notion of unitarity.

Motivated by these ideas and a duality discussed in \cite{Buican:2014hfa,Buican:2017fiq,Buican:2018ddk}, we embarked on a program in \cite{Buican:2017rya} to relate the (typically) logarithmic theories that appear via the correspondence in \cite{Beem:2013sza}\footnote{Note that these theories are sometimes non-unitary but rational. For example, the $(A_1, A_{p-3})$ SCFTs with odd $p\ge5$, which will appear again below, have chiral algebras corresponding to those of the $(2,p)$ Virasoro minimal models.} with a more special set of 2D theories: the unitary rational conformal field theories (RCFTs). These theories, which include the well-known $(m,m+1)$ (where $m\ge3$) Virasoro minimal models as well as various affine Kac-Moody (AKM) theories and even many of the more complicated higher-spin $W$-algebra theories (e.g., see \cite{Bouwknegt:1992wg} for a review), form a very \lq\lq small" subspace in the space of 2D CFTs.

More precisely, in \cite{Buican:2017rya} we studied an infinite class of 4D $\CN=2$ SCFTs called the $D_2[SU(2n+1)]$ theories \cite{Cecotti:2012jx,Cecotti:2013lda}. The corresponding chiral algebras are the logarithmic $\widehat{su(2n+1)}_{-{2n+1\over2}}$ AKM theories (see \cite{Buican:2015ina,Buican:2015hsa} for the $n=1$ case and \cite{Song:2017oew,Xie:2016evu} for $n\ge1$). We then showed that the finite linear combinations of unrefined characters\footnote{By unrefined characters, we just mean the usual sum
\begin{equation}
\chi(q)={\rm Tr}\ q^{L_0}~,
\end{equation}
where $L_0$ is the dilation generator. In particular, we do not turn on any flavor fugacities.
} for admissible\footnote{For an introduction to these types of representations, see \cite{DiFrancesco:1997nk}. Roughly speaking, they are highest weight representations that transform linearly into each other under modular transformations.} representations of $\widehat{su(2n+1)}_{-{2n+1\over2}}$ coincide (up to overall constants) with unrefined characters of the free $\widehat{so(4n(n+1))}_1$ theories. For example, in the case of $n=1$, $D_2[SU(3)]$, we have (up to an overall constant that has been dropped) \cite{Buican:2017rya}
\begin{equation}\label{so8charrel}
\chi_0(q)^{\widehat{su(3)}_{-{3\over2}}}\sim\chi_{1\over2}'^{\widehat{so(8)}_1}(q)~, \ \ \ \chi_{-{1\over2}}'^{\widehat{su(3)}_{-{3\over2}}}(q)\sim\chi_0^{\widehat{so(8)}_1}(q)~,
\end{equation}
where $\chi_0^{\widehat{su(3)}_{-{3\over2}}}(q)$ and $\chi_0^{\widehat{so(8)}_1}(q)$ are the vacuum characters of $\widehat{su(3)}_{-{3\over2}}$ and $\widehat{so(8)}_1$ respectively, $\chi_{1\over2}'^{\widehat{so(8)}_1}(q)$ is the character for a dimension $1/2$ primary of $\widehat{so(8)}_1$ (there are three such primaries, and their unrefined characters are all equal), and $\chi_{-{1\over2}}'^{\widehat{su(3)}_{-{3\over2}}}(q)$ is a finite linear combination of characters for the three primaries with scaling dimension $-1/2$. In these relations, the non-unitary vacuum is mapped to a unitary primary with largest scaling dimension, and a linear combination of the smallest scaling dimension non-unitary primaries is mapped to the unitary vacuum. Given this matching, a main result in \cite{Buican:2017rya} was to find a 4D interpretation of the $\widehat{so(8)}_1$ chiral RCFT (and similarly for $\widehat{so(4n(n+1))}_1$).

While we will briefly return to the discussion of the 4D interpretation in \cite{Buican:2017rya} below, our goals in the present paper are different:
\begin{itemize}
\item{First, we straightforwardly generalize the correspondence in \cite{Buican:2017rya} between logarithmic theories descending from 4D via \cite{Beem:2013sza} and 2D chiral RCFTs to include flavor fugacities as refinements. For simplicity (and because of their more interesting Higgs branches), we will mainly focus on a slightly different class of 4D $\CN=2$ theories, the so-called $(A_1, D_p)$ theories with $p\in\mathbb{Z}_{\rm odd}$.\footnote{We follow the naming conventions of \cite{Cecotti:2010fi}.} However, we will return to the particular theories in \cite{Buican:2017rya} toward the end of our paper.}
\item{Second, we will study the topological quantum field theories (TQFTs)---or, in a more mathematical language, the modular tensor categories (MTCs)\footnote{We will describe the relevant aspects of MTCs in somewhat more detail below. Roughly speaking, MTCs consist of a fusion algebra (in this case a commutative multiplication operation) specified by the action on various simple elements (i.e., elements that are not sums of other elements), a set of matrices, $F$, that implement associativity and satisfy a set of polynomial equations called the \lq\lq pentagon" equations, and a set of braiding matrices, $R$, that, together with the $F$ matrices satisfy the so-called \lq\lq hexagon" equations (e.g., see \cite{Moore:1989vd,etingof2016tensor,bakalov2001lectures}). Moreover, the associated $S$ and $T$ matrices are non-degenerate (and hence the theory is modular).}---underlying the 2D chiral RCFTs, and we will show that these MTCs contain seeds of the IR physics that result from certain Higgs branch RG flows in 4D. In all the examples we will consider, these MTCs are of Chern-Simons type.}
\end{itemize}

At a naive level, one can see an apparently suggestive topological link between the admissible representations of the logarithmic $\widehat{su(3)}_{-{3\over2}}$ chiral CFT and the representations of $\widehat{so(8)}_1$ by constructing the naive fusion rules for the logarithmic theory that follow from applying Verlinde's formula to the modular $S$-matrix for the admissible representations. 
Indeed, labeling the four admissible representations in this theory as $1,a,b,c$ (where $1$ is the vacuum, and $a,b,c$ are dimension $-1/2$ representations), one finds (dropping the trivial $1\otimes x=x$ for $x=1,a,b,c$)
 \begin{equation}\label{fusionD23}
a\otimes a=1~, \ \ \ a\otimes b=-c~, \ \ \ a\otimes c=-b~,\ \ \ b\otimes b=1~,\ \ \ b\otimes c=-a~, \ \ \ c\otimes c=1~.
 \end{equation}
Up to some signs, which reflect the fact that these are not the actual fusion rules of the theory (e.g., see \cite{Gaberdiel:2001ny,Creutzig:2013hma,Ridout:2014yfa,Ridout:2015fga}),\footnote{One issue is that, properly speaking, the admissible modules are not closed under fusion. To find a set of representations that are (conjecturally) closed under fusion one should consider so-called (generalized) \lq\lq relaxed" highest weight modules and their images under spectral flow. We thank Simon~Wood for a discussion on this point.} we find the fusion rules for $\mathbb{Z}_2\times\mathbb{Z}_2$. Still, we might be tempted to interpret these signs as being related in some way to a projective representation of $\mathbb{Z}_2\times\mathbb{Z}_2$. More formally, we may write
\begin{equation}
x\otimes y=\omega(x,y)\cdot z~,
\end{equation}
where $\omega(x,y)\in H^2(\mathbb{Z}_2\times\mathbb{Z}_2, U(1))=\mathbb{Z}_2$ is a 2-cocycle\footnote{In other words, $\omega$ satisfies
\begin{equation}
\omega(h,k) \cdot \omega(g,hk)=\omega(g,h)\cdot \omega(gh,k)~, \ \ \ \omega(1,g)=1~,\ \ \ \forall g,h,k\in\mathbb{Z}_2\times\mathbb{Z}_2~.
\end{equation}} with
\begin{equation}
\omega(a,b)=\omega(b,a)=\omega(a,c)=\omega(c,a)=\omega(b,c)=\omega(c,b)=-1~,
\end{equation}
and all other $\omega=1$. In fact, our $\omega$ is trivial in $H^2(\mathbb{Z}_2\times\mathbb{Z}_2,U(1))$ (i.e., it is a 2-coboundary\footnote{This statement amounts to the fact that $\omega(x,y)=\omega(x)\omega(y)\omega(xy)^{-1}$ with $\omega(1)=1$ and $\omega(a)=\omega(b)=\omega(c)=-1$.}) and so we are naively led to interpret the simple elements as leading to a genuine representation of $\mathbb{Z}_2\times\mathbb{Z}_2$.

While the above analysis is suggestive of a link with $\mathbb{Z}_2\times\mathbb{Z}_2$ fusion rules, we can make a more direct connection by noting that the $\widehat{su(3)}_{-{3\over2}}$ theory is related, at the level of unrefined characters, to the $\widehat{so(8)}_1$ theory via \eqref{so8charrel}. This latter theory has genuine $\mathbb{Z}_2\times\mathbb{Z}_2$ fusion rules! The underlying MTC is just a theory of abelian anyons with a one-form $\mathbb{Z}_2\times\mathbb{Z}_2$ symmetry (see \cite{Gaiotto:2014kfa} for a discussion of one-form symmetries) generated by these anyons.\footnote{At the level of the underlying MTC, one way to describe the full set of results in \cite{Buican:2017rya} is that we associate the two independent MTCs with $\mathbb{Z}_2\times\mathbb{Z}_2$ fusion rules---the $Spin(8)_1$ MTC and the toric code MTC (e.g., see \cite{rowell2009classification,Seiberg:2016rsg} for a discussion of these MTCs)---with the $D_2[SU(2n+1)]$ SCFTs. In particular, if $n(n+1)=0 \ {(\rm mod\ 4)}$, then we associate the toric code MTC with the 4D theory. On the other hand, if $n(n+1)=2 \ {(\rm mod\ 4)}$, then we associate the $Spin(8)_1$ MTC with the theory. Note that the number of admissible representations in $\widehat{su(2n+1)}_{-{2n+1\over2}}$ is $2^{2n}$ \cite{pervse2008vertex}, so this is not, in general, a one-to-one map of admissible representations to simple elements. \label{toricfootnote}}

A different link to abelian anyons appeared recently in the interesting paper \cite{Dedushenko:2018bpp} for the $(A_1, A_3)\simeq(A_1, D_3)$ SCFT and formed some of the motivation for the present paper. There the authors studied new TQFTs coming from AD theories and noted that by \lq\lq flipping the sign" of a simple object in their $(A_1, A_3)$ TQFT, one obtains an MTC with $\mathbb{Z}_3$ fusion rules. In the present context, the naive fusion rules of the admissible representations of the $\widehat{su(2)}_{-{4\over3}}$ chiral algebra associated with the $(A_1, A_3)\simeq(A_1, D_3)$ theory \cite{Buican:2015ina} are
\begin{equation}\label{Z3fusion}
a\otimes b=-1~, \ \ \ a\otimes a=b~, \ \ \ b\otimes b=-a~,
\end{equation}
which, up to two signs, are just $\mathbb{Z}_3$ fusion rules.\footnote{As in the $(A_1, D_4)$ case, it is easy to check that these two signs give rise to a 2-coboundary. This statement is consistent with the fact that $H^2(\mathbb{Z}_3,U(1))=\emptyset$. In particular, by formally taking $a\to-a$ in \eqref{Z3fusion} we recover $\mathbb{Z}_3$ fusion rules.} By solving the hexagon and pentagon equations, it is easy to check that there are two independent unitary MTCs with such $\mathbb{Z}_3$ fusion rules\footnote{There are infinitely many CFTs associated with each of these MTCs since we can take any theory satisfying these fusion rules and tensor in arbitrarily many $(\widehat{e_8})_1$ RCFTs.} (see also \cite{rowell2009classification}): $SU(3)_1$ and $(E_6)_1$. Therefore, it is natural to wonder if there is an associated RCFT whose characters are related to those of the $\widehat{su(2)}_{-{4\over3}}$ theory in a way that parallels the relation in \eqref{so8charrel}.

In fact, an old result of Mukhi and Panda \cite{Mukhi:1989bp} shows the following proportionality of unrefined characters
\begin{equation}
\chi_{0}^{\widehat{su(2)}_{-{4\over3}}}(q)\sim~\chi_{1\over3}'^{\widehat{su(3)}_1}(q)~, \ \ \ \chi_{-{1\over3}}^{'\widehat{su(2)}_{-{4\over3}}}(q)\sim\chi_0^{\widehat{su(3)}_1}(q)~,
\end{equation}
where $\chi_{0}^{\widehat{su(2)}_{-{4\over3}}}(q)$ is the vacuum character of $\widehat{su(2)}_{-{4\over3}}$, $\chi_{-{1\over3}}^{'\widehat{su(2)}_{-{4\over3}}}(q)$ is a finite linear combination of the characters corresponding to the two dimension $-1/3$ representations of $\widehat{su(2)}_{-{4\over3}}$, $\chi_0^{\widehat{su(3)}_1}(q)$ is the vacuum character of $\widehat{su(3)}_1$, and $\chi_{1\over3}'^{\widehat{su(3)}_1}(q)$ is the character of a dimension $1/3$ representation of $\widehat{su(3)}_1$ (there are two such representations, and their unrefined characters are equal). As in \eqref{so8charrel}, the non-unitary vacuum is mapped to a unitary primary with largest scaling dimension, and a linear combination of the smallest scaling dimension non-unitary primaries is mapped to the unitary vacuum. Therefore, we see that the $\widehat{su(3)}_1$ theory is the desired theory related to an MTC with $\mathbb{Z}_3$ fusion rules.

It will be somewhat more useful to think about the $\widehat{su(3)}_1$ characters in terms of the $D$-type modular invariant of $\widehat{su(2)}_4$ \cite{Cappelli:1986hf,Cappelli:1987xt}, which we will denote as $\tilde D_{4}$.\footnote{We add the tilde on top of $\tilde D_4$ to distinguish this $D$ from the one appearing in the related $(A_1, D_3)$ 4D $\CN=2$ SCFT.} This theory can be obtained from $\widehat{su(2)}_4$ by gauging the $\mathbb{Z}_2$ symmetry.\footnote{At the level of the underlying MTC, this procedure corresponds to the evocatively named \lq\lq anyon condensation" \cite{bais2009condensate,neupert2016boson} (see also the recent \cite{Hsin:2018vcg}) and leaves over an MTC with $\mathbb{Z}_3$ fusion rules consisting of anyons having trivial braiding with the anyons generating the $\mathbb{Z}_2$ one-form symmetry in the $SU(2)_4$ MTC.} In particular, one finds \cite{Mukhi:1989bp}
\begin{equation}\label{tildeD4rel}
\chi_{0}^{\widehat{su(2)}_{-{4\over3}}}(q)\sim~\chi_{1\over3}'^{\tilde D_4}(q)=\chi_{2}^{\widehat{su(2)}_4}(q)~, \ \ \ \chi_{-{1\over3}}^{'\widehat{su(2)}_{-{4\over3}}}(q)\sim\chi_0^{\tilde D_4}(q)=\chi_{0}^{\widehat{su(2)}_4}(q)+\chi_{4}^{\widehat{su(2)}_4}(q)~,
\end{equation}
where the $\widehat{su(2)}_4$ characters appearing on the RHS of the above expressions are indexed by an $su(2)$ Dynkin label subscript.

The interpretation in terms of $\widehat{su(2)}_4$ is particularly useful, since now there is a canonical way in which we can try to turn on flavor fugacities in \eqref{tildeD4rel} (the number of fugacities on the LHS and RHS match). As we will see below, there is a discrete subset of fugacities we can turn on so that the characters of $\tilde D_4$ are equal to those of $\widehat{su(2)}_{-{4\over3}}$ up to overall $q$-independent functions. These $q$-independent functions generalize the constants of proportionality we suppressed in writing \eqref{tildeD4rel}. As we will see, a similar story holds for the more general $\widehat{su(2)}_{2(1-p)/p}$ chiral algebras corresponding to the $(A_1, D_p)$ theories with $p\in\mathbb{Z}_{\rm odd}$ and the $\mathbb{Z}_2$ gauging of $\widehat{su(2)}_{2(p-1)}$, $\tilde D_{p+1}$.

The existence of such a matching set of fugacities then motivates us to study RG flows onto the Higgs branch of our $(A_1, D_p)$ theories from the perspective of the related 2D rational chiral algebras and their representations. For $\widehat{su(2)}_{2(1-p)\over p}$, the 2D avatar of the 4D Higgs branch RG flow is just quantum Drinfeld-Sokolov (qDS) reduction \cite{Cordova:2017mhb} (see also \cite{Beem:2014rza,Beem:2013sza} for earlier discussion in other theories).

Instead of performing qDS on the unitary side, we will show that the MTCs underlying our unitary theories \lq\lq know" about certain quantum dimensions (or expectation values for Wilson loop operators) in the non-unitary MTCs related to the IR Higgs branch theories. More precisely, we will argue that these quantum dimensions can be computed after performing a suitable \lq\lq Galois conjugation" \cite{DeBoer:1990em} (see also \cite{Coste:1993af,Harvey:2018rdc}) that takes the unitary RCFT data and makes it non-unitary.

The plan of the paper is as follows. In the next section, we review the $(A_1, D_p)$ theories and place them in a slightly larger context. We also describe the basics of the chiral algebra map in \cite{Beem:2013sza} and how it is applied to these theories. We then review the 2D logarithmic / rational correspondence of characters in \cite{Mukhi:1989bp} and introduce non-trivial flavor fugacities. In the following section we describe how to see topological aspects of the 4D RG flow in the 2D chiral RCFT. Along the way we review relevant aspects of MTCs and Galois conjugation. We conclude with some comments on generalizations of our analysis.

\newsec{The 4D theories and their associated non-unitary chiral algebras}\label{4DRGflow}
Our primary theories of interest are the so-called $(A_1, D_p)$ theories with $p\in\mathbb{Z}_{\rm odd}$. These are 4D SCFTs, sometimes called \lq\lq Argyres-Douglas" (AD) theories, that have $\CN=2$ chiral primaries\footnote{These are superconformal primaries annihilated by all the anti-chiral Poincar\'e supercharges.} of non-integer scaling dimension.\footnote{The $p=3$ case originally appeared in \cite{Argyres:2007cn} generalizing the earlier work in \cite{Argyres:1995jj}.} This property guarantees that they cannot be constructed by standard $\CN=2$ Lagrangians. On the other hand, they can be engineered in at least three other ways: as twisted compactifications of the $A_1$ 6D $(2,0)$ theory \cite{Xie:2012hs,Wang:2015mra}, at the maximally singular points on the Coulomb branches of 4D $\CN=2$ $so(2p)$ Super Yang-Mills (SYM) \cite{Argyres:2012fu}, and as flows from $\CN=1$ Lagrangians via accidental SUSY enhancement \cite{Agarwal:2016pjo} (see also closely related results in \cite{Benvenuti:2017lle,Giacomelli:2017ckh,Agarwal:2018oxb,Maruyoshi:2016aim,Maruyoshi:2016tqk,Agarwal:2017roi,Benvenuti:2017bpg}). For much of the discussion below, the first description will be most intuitive.

To get the $(A_1, D_p)$ theory from the $A_1$ 6D $(2,0)$ theory, we perform a twisted compactification on a twice-punctured $\mathbb{CP}^1$. One puncture is an irregular puncture and one is a \lq\lq full" regular puncture. The \lq\lq full" regular puncture supports the $su(2)$ flavor symmetry of the theory, while the irregular puncture does not have any flavor symmetry associated with it.\footnote{In the Hitchin system description of the theory, the mass parameters are associated with simple poles of the Higgs field. Near the irregular singularity, the Higgs field has more singular behavior and does not include a simple pole.} This picture is useful for us because it gives rise to a natural set of RG flows in 4D: by turning on an expectation value for the moment map operator in the multiplet corresponding to the $su(2)$ flavor symmetry, we can Higgs the regular puncture. In so doing, we go onto the one-quaternionic dimensional Higgs branch of the theory.\footnote{Note that we define the Higgs branch to be the moduli space on which only the $su(2)_R\subset su(2)_R\times u(1)_R$ UV superconformal $R$ symmetry is broken. We do not necessarily mean a branch of moduli space on which there are only free hypermultiplets at generic points.\label{HiggsBr}} Moreover, the remaining irregular singularity supports an $(A_1, A_{p-3})$ theory. There is also a decoupled axion-dilaton hypermultiplet for spontaneous conformal symmetry breaking. As a result, our flow is, up to the decoupled hypermultiplet which we drop\footnote{For further details, see \cite{Xie:2016evu,Song:2017oew,Cordova:2017mhb}.}
\begin{equation}
(A_1, D_p)\to(A_1, A_{p-3})~.
\end{equation} 
The latter $(A_1, A_{p-3})$ SCFTs have no Higgs branches or flavor symmetry themselves and are again strongly interacting Argyres-Douglas theories (the $p=5$ case is the original theory in \cite{Argyres:1995jj}).

In order to unify the results of this paper with our previous work in \cite{Buican:2017rya}, it will be useful to slightly generalize the theories we are studying and consider the $(A_{N-1}^N[p-N],F)$ SCFTs with $p$ and $N$ co-prime integers (e.g., see \cite{Xie:2016evu,Song:2017oew}). The above discussion was for the case of $N=2$. In particular
\begin{equation}\label{theoryID}
(A_1, D_p)\sim (A_{1}^2[p-2],F)~, \ \ \ (A_1, A_{p-3})\sim A_{1}^2[p-2]~.
\end{equation}
However, the pattern for general $N$ is similar: these theories are compactifications of the $A_{N-1}$ 6D $(2,0)$ theory on a $\mathbb{CP}^1$ with an irregular and \lq\lq full" regular puncture. This latter puncture supports an $su(N)$ flavor symmetry with level
\begin{equation}
k^{4d}_{su(N)}={2N(p-1)\over p}~,
\end{equation}
while the irregular puncture does not have any flavor symmetry associated with it. We can again fully Higgs the regular puncture and obtain the following RG flow (where again we drop decoupled free hypermultiplets) to a theory with just an irregular puncture
\begin{equation}\label{RGflowH}
(A_{N-1}^N[p-N],F)\to A_{N-1}^N[p-N]~.
\end{equation}
The $A_{N-1}^N[p-N]$ theory is again an interacting SCFT only if $p>N$.\footnote{This statement is not an if and only if: the theory with $p=3$ and $N=2$ is trivial.} The central charges of the theories appearing in the above flow are \cite{Xie:2016evu,Song:2017oew}
\begin{equation}\label{central}
c_{(A_{N-1}^N[p-N],F)}={p-1\over12}(N^2-1)~, \ \ \ c_{(A_{N-1}^N[p-N],F)}={(N-1)(p-1-N)(p+(p-N)N)\over12 p}~.
\end{equation}

As a final point, note that we can consider more general RG flows than the ones in \eqref{RGflowH}. Indeed, we can consider RG flows in which we only partially Higgs the regular puncture (and break the associated global symmetry group to some more general subgroup). In these cases, we can have more complicated theories in the IR. These flows will play a role when we return to discuss the theories in \cite{Buican:2017rya}.

In the next section, we will consider the 2D chiral algebras in the sense of \cite{Beem:2013sza} that correspond to the two endpoints in \eqref{RGflowH}.

\subsec{The non-unitary chiral algebras}
The authors in \cite{Beem:2013sza} found a very general connection between certain operators---called Schur operators---that sit in short multiplets of any local 4D $\CN=2$ SCFT and meromorphic currents that generate 2D chiral algebras. In this section we will not try to give a thorough review of this construction. Instead, we will conduct a quick review and highlight relevant aspects of \cite{Beem:2013sza} for the theories at hand. We refer the interested reader to \cite{Beem:2013sza} for further details.

Two particularly important types of multiplets containing Schur operators are the stress-tensor multiplet, $\widehat{\CC}_{0(0,0)}$, and the flavor current multiplet, $\widehat{B}_1$.\footnote{Here we use the nomenclature of \cite{Dolan:2002zh} (see also the earlier work in \cite{Dobrev:1985qv}). } In the case of $\widehat{\CC}_{0(0,0)}$, the Schur operator is the highest Lorentz and $R$-weight component of the $su(2)_R$ symmetry current, $J^{ij}_{\alpha\dot\alpha}|_{\rm hw}=J_{+\dot+}^{11}$ (recall that, as reviewed in footnote \ref{HiggsBr}, 4D $\CN=2$ SCFTs have an $su(2)_R\times u(1)_R$ superconformal $R$ symmetry). In our notation, $\alpha\in\left\{\pm\right\},\dot\alpha\in\left\{\dot\pm\right\}$ are chiral and anti-chiral Lorentz spinor indices respectively, and $i,j\in\left\{1,2\right\}$ are $su(2)_R$ spin half indices. In the case of the flavor current multiplet, the Schur operator is the $su(2)_R$ highest-weight component of the moment map superconformal primary (i.e., the holomorphic moment map), $J^{ij}|_{\rm hw}=\mu$. In writing the holomorphic moment map, we have suppressed adjoint indices for the flavor symmetry corresponding to the partner flavor current.

The most important feature of a Schur operator, $\CO$ (here we suppress $su(2)_R$ and Lorentz indices), is that it satisfies
\begin{equation}\label{Schurcond}
\left\{\mathbbmtt{Q},\mathcal{O}(0)\right]=0~, \ \ \ \mathcal{O}(0)\ne\left\{\mathbbmtt{Q},\mathcal{O}'(0)\right]~, \ \ \ \mathbbmtt{Q}=S_1^{-}-\tilde Q_{2\dot-}~,
\end{equation}
where, $\tilde Q_{2\dot-}$ is a Poincar\'e supercharge, and $S_1^-$ is a special supercharge of the 4D $\CN=2$ superconformal algebra. In other words, we see from \eqref{Schurcond} that these operators form non-trivial $\mathbbmtt{Q}$ cohomology classes (other operators in the theory do not).

Given these facts, the main insight of \cite{Beem:2013sza} is that one can obtain an interesting algebra of operators by placing the Schur operators in a plane, $P=\mathbb{C}\subset\mathbb{R}^4$, and twisting the anti-holomorphic $sl(2,\mathbb{R})$ conformal subgroup in the plane by $su(2)_R$ transformations. These twists are done in such a way that they render transformations in $\bar z$ $\mathbbmtt{Q}$-exact. Therefore, by restricting to $\mathbbmtt{Q}$ cohomology classes of Schur operators, we find a map to a set of meromorphic currents in $P$. In particular, we have the following natural maps
\begin{equation}
\chi[J_{+\dot+}^{11}]=T~, \ \ \chi[\mu]=J~,
\end{equation}
where $T$ and $J$ are the holomorphic 2D stress tensor and 2D AKM current respectively. As a result, any local 4D $\CN=2$ SCFT has a 2D chiral algebra that contains a Virasoro sub-algebra. If the theory has a locally realized flavor symmetry, then the related chiral algebra has an AKM sub-algebra. The structure of current-current OPEs implies that
\begin{equation}
c_{2d}=-12c_{4d}~, \ \ \ k_{2d}=-{1\over2}k_{4d}~.
\end{equation}
In particular, if the 4D theory is unitary, the 2D one is not.

One additional useful piece of data captured by the above correspondence is the so-called Schur index \cite{Gadde:2011ik,Gadde:2011uv} of the 4D theory
\begin{equation}
\CI_S(q,x_i)=q^{c_{4d}\over2}{\rm Tr}_{\CH}(-1)^Fq^{E-R}\prod_ix_i^{f_i}~.
\end{equation}
This is a refined Witten index that counts the Schur operators weighted by fermion number, $F$, flavor fugacities, $x_i\in u(1)$, with charges $f_i$, and a superconformal fugacity, $q$, satisfying $|q|<1$, weighted by the difference of the 4D scaling dimension, $E$, and the $su(2)_R$ weight, $R$. Under the correspondence in \cite{Beem:2013sza}, this index is naturally mapped to the torus partition function of the 2D chiral algebra
\begin{equation}\label{indtorus}
\CI_S(q,x_i)=Z(-1,q,x_i)~, \ \ \ Z(y,q,x_i)=q^{-{c_{2d}\over24}}{\rm Tr}\ y^{M^{\perp}}q^{L_0}x_i^{f_i}~,
\end{equation}
where $M^{\perp}$ generates rotations in the plane normal to $P$. This character forms part of a representation of the modular group \cite{Beem:2017ooy}. The 4D interpretation of the modular partners of the vacuum character are as indices of operators living on certain $\CN=(2,2)$-preserving surface defects (e.g., see \cite{Cordova:2017mhb,Nishinaka:2018zwq} for some examples).

By applying this map to the $(A_{N_1}^N[p-N],F)$ and $A^N_{N-1}[p-N]$ SCFTs, the authors of \cite{Xie:2016evu,Song:2017oew,Cordova:2015nma} found
\begin{equation}\label{chiralgs}
\chi[(A_{N-1}^N[p-N],F)]=\widehat{su(N)}_{N{1-p\over p}}~, \ \ \ \chi[A^N_{N-1}[p-N]]=W_{N-1}(N,p)~,
\end{equation}
where $W_{N-1}(N,p)$ is the chiral algebra of the $A_{N-1}$ $W$-algebra minimal model. In particular, for the case of $N=2$, $W_1(2,p)$ is just the algebra of the $(2,p)$ Virasoro minimal model. Interestingly, the indices for the UV theories take a particularly simple form \cite{Xie:2016evu,Song:2017oew}\footnote{See \cite{Buican:2015ina,Cordova:2015nma} for earlier work on subsets of these theories (and also closely related work in \cite{Buican:2017uka,Song:2015wta,Buican:2015tda}).}
\begin{equation}
\CI_{S,(A_{N-1}^N[p-N],F)}={\rm P.E.}\left({q-q^p\over(1-q)(1-q^p)}\chi_{\rm adj}\right)~, \ \ \ P.E.(f(x_i))\equiv{\rm Exp}\left(\sum_{n=1}^{\infty}{1\over n}f(x_i^n)\right)~,
\end{equation}
where $\chi_{\rm adj}$ is an adjoint character for $su(N)$. Indeed, this result has been mathematically proven (assuming the correspondence in \eqref{chiralgs}) for so-called \lq\lq boundary admissible" theories in 2D \cite{kac2017remark,Creutzig:2017qyf} (this class of theories includes the $\widehat{su(N)}_{N{1-p\over p}}$ theories).

\newsec{From logarithmic theories to RCFT}\label{logtorat}
For much of the remainder of the paper, we will be concerned with the case of $N=2$. In particular, the relevant logarithmic chiral algebras will be
\begin{equation}
\chi[(A_1, D_p)]=\chi[(A_1^2[p-2],F)]=\widehat{su(2)}_{2(1-p)\over p}~,
\end{equation}
with positive $p\in\mathbb{Z}_{{\rm odd}}$. As we briefly mentioned in the introduction, unrefined characters for these chiral algebras and their admissible representations were studied in \cite{Mukhi:1989bp}, where the authors found an interesting connection with unrefined characters of the rational and unitary $\widehat{su(2)}_{2(p-1)}$ algebras and representations.

These latter AKM algebras have $2p-1$ primaries, $\Phi_{\ell}$ (here $\ell\in\left\{0,1,\cdots,2(p-1)\right\}$ is an $su(2)$ Dynkin label), with conformal dimensions
\begin{equation}\label{su2Udims}
h(\Phi_{\ell})={\ell(\ell+2)\over8p}~.
\end{equation}
The $\Phi_{\ell}$ satisfy the following fusion algebra \cite{Gepner:1986wi}
\begin{equation}
\Phi_{\ell_1}\otimes\Phi_{\ell_2}=\sum_{\ell=|\ell_1-\ell_2|}^{{\rm min}(|\ell_1+\ell_2|,4(p-1)-\ell_1-\ell_2)}\Phi_{\ell}~.
\end{equation}
Note that the $\Phi_{2(p-1)}$ field satisfies $\mathbb{Z}_2$ fusion rules, $\Phi_{2(p-1)}\otimes\Phi_{2(p-1)}=\Phi_0$, and is associated with the non-trivial element of $\mathbb{Z}_2$ (here $\Phi_0=1$). More precisely, the associated topological defect (see \cite{Buican:2017rxc} for a recent discussion) implements the $\mathbb{Z}_2$  symmetry of the $\widehat{su(2)}_{2(p-1)}$ CFT. Equivalently, we can think of $\Phi_{2(p-1)}$ as corresponding to the abelian anyon in the related Chern-Simons theory (e.g., see the classic works \cite{Moore:1989vd,Witten:1988hf}) that generates the $\mathbb{Z}_2$ one-form symmetry.

As discussed in the introduction, the particular theories that the authors studied in \cite{Mukhi:1989bp} are actually $\mathbb{Z}_2$ oribfolds of the $\widehat{su(2)}_{2(p-1)}$ theories. We label these theories as $\tilde D_{p+1}$ (or $\widehat{psu(2)}_{2(p-1)}\simeq\widehat{so(3)}_{2(p-1)}$), and they are the chiral parts of the $D$-type modular invariants in \cite{Cappelli:1986hf,Cappelli:1987xt}. Gauging the $\mathbb{Z}_2$ symmetry projects out fields that are not invariant under the action of the corresponding topological defect, i.e. those fields satisfying
\begin{equation}\label{su2S}
{S_{2(p-1),\ell}\over S_{1,\ell}}\ne 1~, \ \ \ S_{\ell_1,\ell_2}={1\over\sqrt{p}}\sin\left[{(\ell_1+1)(\ell_2+1)\pi\over2p}\right]~,
\end{equation}
where $S_{\ell_1,\ell_2}$ is the modular $S$-matrix of $\widehat{su(2)}_{2(p-1)}$ \cite{Gepner:1986wi}. This projection immediately eliminates the (half-integer spin) odd $\ell$ fields. Next, one organizes primaries into representations of a larger chiral algebra by associating each representation with the orbit under fusion with $\Phi_{2(p-1)}$ and treating fixed points separately. There is one fixed point under this fusion since $\Phi_{2(p-1)}\otimes\Phi_{p-1}=\Phi_{p-1}$, and so one associates $|\mathbb{Z}_2|=2$ representations of the enlarged chiral algebra with this representation, $\Phi_{D, p-1}^i$ where $i=1,2$. In other words, our theory after $\mathbb{Z}_2$ gauging is just given in terms of the following representations of the original theory\footnote{In the condensed matter literature, the corresponding Chern-Simons MTC is said to have undergone anyonic condensation.}
\begin{equation}\label{Dreps}
\Phi_{D,\ell}=\Phi_{\ell}\oplus\Phi_{2(p-1)-\ell}~, \ \ \ \ell\in\{0,2,4,\cdots,p-3\}~, \ \ \ \Phi_{D,p-1}^i=\Phi_{p-1}^i~.
\end{equation}
As a result, there are $(p+3)/2$ representations and $(p+1)/2$ independent characters since the characters for $\Phi_{p-1}^i$ are equal
\begin{equation}
\chi_{D,p-1,1}(q)=\chi_{D,p-1,2}(q)=\chi_{p-1}^{\widehat{su(2)}_{2(p-1)}}(q)~.
\end{equation}

On the other hand, the $\widehat{su(2)}_{2(1-p)\over p}$ algebra has $p$ admissible representations, $\hat\Phi_j$, with $j=0,1,2,\cdots, p-1$ and scaling dimensions
\begin{equation}
h(\hat\Phi_j)=-{j\over2}\left({p-j\over p}\right)~.
\end{equation}
In the limit that we turn off flavor fugacities, all the corresponding characters except the vacuum character are divergent. However, the following linear combinations of non-unitary characters are finite
\begin{equation}
\chi_{-,0}(q)\equiv\chi_0(q)~, \ \ \ \chi_{-,j}\equiv\chi_j(q)-\chi_{p-j}(q)~, \ \ \ j=1,2,\cdots,{p-1\over2}~.
\end{equation}
Clearly, there are $(p+1)/2$ such characters, which matches the number of independent characters in the unitary case.

Given these sets of characters on the unitary and non-unitary sides, one of the main results of \cite{Mukhi:1989bp} is that, up to overall constants, we have
\begin{equation}\label{MPrel}
\chi_{D,p-1-2j}(q)\sim\chi^{\widehat{su(2)}_{2(1-p)\over p}}_{-,j}(q)~.
\end{equation}
In other words, the unrefined character for the Dynkin label $p-1-2j$ primary of $\widehat{su(2)}_{2(p-1)}$ is proportional to the unrefined character of the $j^{\rm th}$ non-unitary primary.

In the next subsection we will briefly expand on this result and introduce discrete flavor fugacities for $su(2)$. This matching then motivates us to study RG flows onto the 4D Higgs branch from the perspective of the unitary 2D theory.

\subsec{Flavoring the correspondence}
Let us consider turning on the $su(2)$ flavor fugacity, $y$, in the above correspondence. For simplicity, we will limit ourselves to $j=0$. This case is the most immediately interesting from the 4D perspective since the $j=0$ non-unitary character is the 4D Schur index of the $(A_1, D_p)$ SCFT (see \eqref{indtorus} and \eqref{chiralgs}).

For generic $y\in u(1)$, the refined characters are related in relatively complicated ways. However, it is straightforward to show that the two characters agree up to a $q$-independent function of $y$ when $y$ is a $(p+1)^{\rm st}$ root of unity.\footnote{This statement holds somewhat more generally.} More precisely, we have
\begin{equation}\label{rel1}
\chi_{D,p-1}(q,y)=\chi_{su(2),p-1}(y)\cdot\chi_0^{\widehat{su(2)}_{2(1-p)\over p}}(q,y)~, \ \ \ y=y_k=e^{2\pi i k\over p+1}~,
\end{equation}
where $\chi_{su(2),p-1}(y)=\sum_{i=-{p-1\over2}}^{p-1\over2}y^i$ is a spin $(p-1)/2$ character of $su(2)$. At the discrete points $y_k=e^{2\pi i k\over p+1}\ne1$, we have $\chi_{su(2),p-1}(y_k)=(-1)^{1+k}$ while $\chi_{su(2),p-1}(y_0)\equiv\chi_{su(2),p-1}(1)=p$, and so
\begin{equation}\label{flavoredchar}
\chi_{D,p-1}(q,y_k)=
\begin{cases}
(-1)^{1+k}\cdot\chi_{-,0}^{\widehat{su(2)}_{2(1-p)\over p}}(q,y_k)~, &\text{if}\ 1\le k\le p\\
p\cdot\chi_{-,0}^{\widehat{su(2)}_{2(1-p)\over p}}(q,y_k)~, &\text{if}\ k=0~.
\end{cases}
\end{equation}

To prove \eqref{flavoredchar}, we start by writing the explicit forms of the two characters. For the non-unitary vacuum character, we have \cite{DiFrancesco:1997nk,Mukhi:1989bp,Kac:1988qc}
\begin{equation}\label{logchar}
 \chi_{-,0}^{\widehat{su(2)}_{2(1-p)\over p}}\left(q,y\right) =  \frac{\Theta_{p}^{(2p)}(q,y^{1\over p}) - \Theta_{-p}^{(2p)}(q,y^{1\over p}) }{\Theta_{1}^{(2)}(q,y) - \Theta_{-1}^{(2)}(q,y)} \ , 
\end{equation}
where $y = e^{-2\pi i z}$, and
\begin{equation}
\Theta_{j}^{(k)}(q,x) = x^{\frac{j}{2}} q^{\frac{j^{2}}{4k}} \sum_{n \in \mathbb{Z}} x^{kn} q^{kn^{2} + nj} \ .
\end{equation}
Similarly, the rational character is given by \cite{DiFrancesco:1997nk,Mukhi:1989bp,Kac:1988qc}
\begin{equation}\label{ratchar}
  \chi_{(p-1)_D}\left(q,y\right) =  \frac{\Theta_{p}^{(2p)}(q,y) - \Theta_{-p}^{(2p)}(q,y) }{\Theta_{1}^{(2)}(q,y) - \Theta_{-1}^{(2)}(q,y)} \ . 
\end{equation}
In particular, the denominators in \eqref{logchar} and \eqref{ratchar} agree. The numerators are closely related as well. Indeed, the numerator of \eqref{logchar} is
\begin{equation}
\Theta_{p}^{(2p)}(q,y^{1\over p}) - \Theta_{-p}^{(2p)}(q,y^{1\over p}) = \sum_{n \in \mathbb{Z}} \left( y^{2n+\frac{1}{2}} - y^{-\left(2n +\frac{1}{2}\right)} \right) q^{{p\over2}\left(2n+\frac{1}{2}\right)^{2}} ~,
\end{equation}
while the numerator of \eqref{ratchar} is
\begin{equation}
\Theta_{p}^{(2p)}(q,y) - \Theta_{-p}^{(2p)}(q,y)=\sum_{n \in \mathbb{Z}} \left( y^{p\left(2n+\frac{1}{2}\right)} - y^{-p\left(2n +\frac{1}{2}\right)} \right) q^{{p\over2}\left(2n+\frac{1}{2}\right)^{2}} ~.
\end{equation}
Asking that the characters be proportional to each other up to a function that is independent of $q$ requires that we choose values of $y$ such that the ratio
\begin{equation}\label{rdef}
r(y,n)={y^{p\left(2n+\frac{1}{2}\right)} - y^{-p\left(2n +\frac{1}{2}\right)}\over y^{2n+\frac{1}{2}} - y^{-\left(2n +\frac{1}{2}\right)}}~,
\end{equation}
is independent of $n$. This condition is satisfied when $y=y_k$. To verify this statement, first suppose $k\ne0$. Then, the numerator and denominator in \eqref{rdef} do not vanish, and
\begin{eqnarray}
r(y,n)={\sin \left( 2 \pi \frac{(2n + \frac{1}{2}) k p}{p+1}\right)\over \sin \left( 2\pi \frac{(2n + \frac{1}{2}) k}{p+1}\right)}={\sin \left( 2 \pi k \left(2n + \frac{1}{2}\right) - 2 \pi \frac{\left(2n + \frac{1}{2}\right)k}{p+1} \right)\over \sin \left( 2\pi \frac{(2n + \frac{1}{2}) k}{p+1}\right)}=(-1)^{1+k}~.
\end{eqnarray}
If $k=0$, then we find $r(y_0,n)\equiv\lim_{y\to1}r(y,n)=p$ as desired (the characters themselves do not degenerate, because the denominators in \eqref{logchar} and \eqref{ratchar} also vanish at the same order).

The simple relations in \eqref{rel1} and \eqref{flavoredchar} for $y\ne1$ suggest that $\tilde D_{p+1}$ should know something about the Higgs branch of the $(A_1, D_p)$ SCFT. Indeed, from the 4D perspective, we can learn about the index of the Higgs branch theory by considering poles in the flavor fugacity, $y$ \cite{Gaiotto:2012xa}.\footnote{In fact, we will see that for general $p$ we most directly learn something about the 4D theory in the presence of a surface defect.}

\newsec{MTCs and the RG flow}
In section \ref{4DRGflow}, we saw that there were interesting RG flows emanating from the $(A_1, D_p)$ fixed points that take us onto their Higgs branches
\begin{equation}\label{RGflowA}
(A_1, D_p)\to (A_1, A_{p-3})~.
\end{equation}
In writing \eqref{RGflowA}, we have dropped a decoupled hypermultiplet containing goldstone bosons and their superpartners. Since moving onto the Higgs branch requires breaking flavor symmetry, and since we showed in the previous section that $\tilde D_{p+1}$ knows about certain (discretely) flavored observables in the $(A_1, D_p)$ SCFT, one might be tempted to guess that we can learn about the 4D Higgs branch using the 2D chiral RCFT.

We will see that this intuition is indeed correct, although perhaps not in the most obvious way one might first imagine. Indeed, as a first guess, one might try to perform qDS reduction on the $\tilde D_{p+1}$ theory, since this reduction applied to the 2D chiral algebra of the $(A_1, D_p)$ theory gives the 2D chiral algebra of the $(A_1, A_{p-3})$ theory (e.g., see the discussion in \cite{Cordova:2017mhb}). Instead, we will describe a simpler connection.

The idea is to consider some of the most basic data in the Chern-Simons theories underlying the $\tilde D_{p+1}$ theories: the $S^3$ expectation values of Wilson loops, $W_{D,p-1}^i$, corresponding to the highest-spin primaries, $\Phi_{D,p-1}^i$ (we will see that the answer does not depend on $i$)
\begin{equation}\label{Wilson}
\langle W_{D,p-1}^i\rangle={S_{0,(p-1)_i}\over S_{0,0}}~,
\end{equation}
where $S_{a,b}$ is the modular $S$-matrix for $\tilde D_{p+1}$. We will show that this data can be related---via Galois conjugation---to the expectation value of a Wilson loop in the TQFT underlying the chiral part of the $(2,p)$ Virasoro minimal model (i.e., the 2D theory for the IR $(A_1, A_{p-3})$ SCFT in the sense of \cite{Beem:2013sza}). More precisely, the expectation value in question is for the Wilson loop corresponding to the lowest scaling dimension primary, $\phi_{(1,(p-1)/2)}$.

In order to understand these statements and their implications, we review basic aspects of MTCs and Galois conjugation in the next subsection. We then move on to discuss the action of the RG flow on \eqref{Wilson}.

\subsec{MTC / TQFT basics}
Roughly speaking, to the representations of any 2D rational chiral algebra, we can associate a corresponding MTC (or 3D TQFT depending on one's preference) \cite{Moore:1989vd,Witten:1988hf}. In our cases of interest, these MTCs are of Chern-Simons type. The general data that defines an MTC is a set of simple objects with corresponding commutative fusion rules,  a set of $F$ matrices that implement associativity and satisfy \lq\lq pentagon" equations, and a set of braiding or $R$ matrices that satisfy, together with the $F$ matrices, the so-called \lq\lq hexagon" equations \cite{Moore:1989vd}. The MTC is modular because it has associated with it non-degenerate $S$ and $T$ matrices. Since our MTCs arise from representations of 2D rational chiral algebras, the resulting simple objects are in one-to-one correspondence with the representations of these chiral algebras. In a Chern-Simons theory, one thinks of these simple objects as tracing out Wilson lines in some representation of the gauge group. As can be seen by studying their braiding properties, these objects are generally anyonic.

For us in what follows, the most important data in the MTC will be the $S$ and $T$ matrices. The $T$ matrices we use consist of the twists (unnormalized by the standard RCFT prefactor, $e^{-2\pi i{c\over24}}$)
\begin{equation}\label{Tmatdef}
T_{i,j}=\delta_{ij}\theta_i=\delta_{ij}e^{2\pi i h_i}~,
\end{equation}
where $h_i$ is the conformal dimension of the corresponding primary, $\Phi_i$, of the 2D rational chiral algebra. Another important piece of data for us is the set of quantum dimensions
\begin{equation}
d_i={S_{0i}\over S_{00}}~,
\end{equation}
corresponding to the expectation value of a Wilson loop of type $i$ on $S^3$. Since our starting point is unitary, we have
\begin{equation}\label{dimbound}
d_i\ge1~,
\end{equation}
where $d_i=1$ if and only if the corresponding anyon is abelian, i.e. if there exists $\bar i$ (which may or may not satisfy $i=\bar i$) such that
\begin{equation}
i\otimes\bar i =1~.
\end{equation}
Note that this fusion rule corresponds to the RCFT fusion $\phi_{i}\otimes\phi_{\bar i}=\phi_{0}$, where the $\phi_a$ are RCFT primaries ($\phi_{0}$ is the identity). The proof of this statement follows from the fact that $d_1=1$, $d_i=d_{\bar i}$, and the fact that the quantum dimensions satisfy the fusion rules of the theory \cite{DiFrancesco:1997nk,kitaev2006anyons}
\begin{equation}
d_jd_k=\sum_kN^{\ell}_{jk}d_{\ell}~,
\end{equation}
where the integers $N^{\ell}_{jk}\ge0$ are the fusion multiplicities. As a result, the $i$ anyon generates (part of) the abelian one-form symmetry of the theory (and $\bar i$ is $i$'s \lq\lq inverse"). We call such an anyon an \lq\lq abelian" anyon to distinguish it from the anyons, $a$, with $d_a>1$, whose fusion rules are not those of a group ($a\times\bar a$ will involve at least two non-trivial fusion channels).

\subsubsection{Galois conjugation}
Given an MTC, we may define various natural actions on it. One particularly important action is that of Galois conjugation. While the precise action of Galois conjugation at the level of the full MTC is subtle,\footnote{One reason is  that some of the data in the $F$ and $R$ matrices depends on certain gauge choices.} a Galois action at the level of the generalized quantum dimensions\footnote{These include not only the $d_i={S_{i0}\over S_{00}}$ but also the ${S_{ij}\over S_{0j}}$ with $j\ne0$.} is simpler to describe \cite{deBoer:1993iz,Coste:1993af}.

The main point is that the quantum dimensions can be thought of as taking values in some \lq\lq cyclotomic" field, $\mathbb{Q}(\xi)$, for $\xi=e^{2\pi i\over k}$, which consists of appending $k$th roots of unity to the rational numbers, $\mathbb{Q}$.\footnote{A similar story holds for the modular $S$ and $T$ data, although the cyclotomic field is, in general, different \cite{Coste:1993af}. We will comment further on this fact below.} The cyclotomic field admits the action of a Galois group, $G=\mathbb{Z}_k^{\times}$, consisting of the multiplicative units between $1$ and $k$ (e.g., $\mathbb{Z}_4^{\times}=\left\{1,3\right\}$). The action of $G$ is simple to describe: it leaves the base field (i.e., the rational numbers) invariant and acts non-trivially on $\xi$ as
\begin{equation}\label{galoisAct}
\xi\to\xi^p~, \ \ \ p\in\mathbb{Z}_k^{\times}~,
\end{equation}
for $p$ and $k$ co-prime. In general, Galois conjugation takes unitary theories to non-unitary ones (although there are exceptions). The mosts basic example being the Galois action that takes the Lee-Yang MTC to $(G_2)_1$, $(F_4)_1$, and the complex conjugate of Lee-Yang. We will return to this example shortly. Note that in the non-unitary conjugates of a unitary theory, the quantum dimension bound in \eqref{dimbound} is typically violated. 

Before proceeding, let us emphasize that the examples of Galois group we discuss here can be naturally related to one that acts on the full $S$ and $T$ matrices in an RCFT \cite{Bantay:2001ni,Coste:1993af} by a surjective restriction (and similarly for the natural Galois action descending from the quantum group structure underlying the MTC).

\subsec{Galois action, RG flows, and quantum dimensions}
In this section, we study the action of a Galois group on some of the data underlying the $\tilde D_{p+1}$ theory. We start with the special cases of $p=3$ and $p=5$ before discussing the general case. As we will see, some additional interesting phenomena occur for $p=3,5$.

To that end, consider the case of $p=3$. As discussed in section \ref{logtorat}, the resulting $\tilde D_4$ theory is a theory with abelian fusion rules. Indeed, after anyon condensation in $SU(2)_{4}$, the resulting Chern-Simons theory has abelian anyons and $\mathbb{Z}_3$ fusion rules. From \eqref{RGflowA}, we see that the resulting 4D IR theory is the trivial $(A_1, A_0)$ theory.\footnote{One can also see from \eqref{central} that the corresponding central charge with $p=3$ and $N=2$ vanishes (in the discussion below \eqref{chiralgs}, this is because the IR chiral algebra is for the trivial $(2,3)$ Virasoro minimal model).} Later we will see that this phenomenon appears in other examples as well: when the UV theory consists of abelian anyons, the 4D Higgs branch theory is either trivial (after removing the decoupled hypermultiplet of spontaneous symmetry breaking) or free (at generic points). This statement is also consistent with the matching of quantum dimensions alluded to in the introduction
\begin{equation}\label{p3S}
\langle W_{D,2}^i\rangle=d_{2_i}={S_{0,2_i}\over S_{0,0}}=1~, \ \ \ S^D={1\over\sqrt{3}}
  \begin{pmatrix}
    1 & 1 & 1 \\
    1 & \omega & \omega^2 \\
    1 & \omega^2 & \omega \\
  \end{pmatrix}~.
\end{equation}
In writing the $S$-matrix, we have taken the second (third) row / column to correspond to $2_1$ ($2_2$). These rows and columns correspond to the anyons that generate $\mathbb{Z}_3$. Indeed, as we explained in the previous subsection, anyons whose fusion rules are abelian have quantum dimension one. The IR theory is trivial (after considering the 2D theory related to the 4D theory we get by dropping the Goldstone multiplet) and so the only IR field is the vacuum, $\phi_{(1,1)}$, with quantum dimension one.

Next let us consider the case of $p=5$, i.e., $\tilde D_6$. The corresponding modular $S$-matrix is\footnote{The modular $S$-matrix can be derived from the one for $\widehat{su(2)}_8$ as follows. First, note that the primaries of the $\tilde D_6$ chiral algebra are fixed in terms of the $\widehat{su(2)}_8$ primaries as in \eqref{Dreps}. This observation fixes the first three rows / columns in the modular $S$-matrix in terms of the entries in the $S$-matrix in \eqref{su2S}. The remaining two rows and columns (i.e., for the two $\Phi_{4}^i$ primaries) can be fixed by demanding symmetry of the $S$-matrix, reality of the first row (and column), unitarity, and the $sl(2,\mathbb{Z})$ conditions $S^2=(ST)^3$ and $S^4=1$.}
\begin{equation}
S^D=\begin{pmatrix}
    {1\over10}\left(5-\sqrt{5}\right) & {1\over10}\left(5+\sqrt{5}\right) & {1\over\sqrt{5}} & {1\over\sqrt{5}} \\
    {1\over10}\left(5+\sqrt{5}\right) & {1\over10}\left(5-\sqrt{5}\right) & -{1\over\sqrt{5}} & -{1\over\sqrt{5}} \\
    {1\over\sqrt{5}} & -{1\over\sqrt{5}} & -{1\over10}\left(5-\sqrt{5}\right) & {1\over10}\left(5+\sqrt{5}\right)\\ {1\over\sqrt{5}} & -{1\over\sqrt{5}} & {1\over10}\left(5+\sqrt{5}\right) & -{1\over10}\left(5-\sqrt{5}\right)\\
\end{pmatrix}~.
\end{equation}
Now, using the Verlinde formula 
\begin{equation}
N^{\nu}_{\lambda\mu}=\sum_{\sigma}{S_{\lambda,\sigma}S_{\mu,\sigma}S^*_{\sigma,\nu}\over S_{0_D,\sigma}}~,
\end{equation}
we find that
\begin{equation}
\Phi^i_{D,4}\times\Phi^i_{D,4}=\Phi_{D,0}+\Phi^i_{D,4}~.
\end{equation}
In particular, we see that $\left\{\Phi_{D,0},\Phi^i_{D,4}\right\}$ are closed fusion subcategories (one two-element subcategory for each value of $i$; without loss of generality, we will drop $i$ from now on). Moreover, their fusion rules are the so-called \lq\lq Fibonnaci" fusion rules (e.g., see \cite{wang2010topological} for a review) shared by the Lee-Yang, conjugate Lee-Yang, $(G_2)_1$, and $(F_{4})_1$ fusion categories. In our case, after normalizing the sub-$S$-matrix for $\left\{\Phi_{0_D},\Phi_{4_D}\right\}$, we obtain
\begin{eqnarray}
S&=& {1\over\sqrt{\xi^{-1}+3+\xi}}\begin{pmatrix}
     1& \xi^{-1}+1+\xi \\
      \xi^{-1}+1+\xi&-1  \\
  \end{pmatrix}~, \ \ \ d_{\Phi_{D,0}}=1~, \ \ \ d_{\Phi_{D,4}^i}=\xi^{-1}+1+\xi\ \ \ ~, \cr T &=& {\rm diag}(1,\xi^3) ~, \ \ \ \xi=e^{2\pi i\over5}~.
\end{eqnarray}
These are the $S$ and $T$ matrices for the $(F_4)_1$ MTC \cite{rowell2009classification}. Using Galois conjugation as in \eqref{galoisAct} at the level of the $S$ and $T$ matrices, we can transform the above data into the data for Lee-Yang. More precisely, if we Galois conjugate by the element $2\in\mathbb{Z}_5^{\times}$, we obtain\footnote{Note that the Galois group studied in \cite{Bantay:2001ni,Harvey:2018rdc} is $\mathbb{Z}_{60}^{\times}$. The reason for this difference is that the authors of these latter works consider the CFT-normalized $T$ matrix (i.e., with the $e^{-{2\pi ic\over24}}$ prefactor). There is no inconsistency in using these two different groups since we have an appropriate surjective restriction $\mathbb{Z}_{60}^{\times}\to\mathbb{Z}_5^{\times}$.}
\begin{eqnarray}
d^{\overline{LY}}_{\phi_{(1,1)}}=1~, \ \ \ d^{\overline{LY}}_{\phi_{(1,2)}}=\xi^{-2}+1+\xi^2~, \ \ \  T^{\overline{LY}} = {\rm diag}(1,\xi^6) ~,
\end{eqnarray}
which is the complex conjugate of the Lee-Yang category. On the other hand, if we conjugate by the element $3\in\mathbb{Z}_5^{\times}$, we obtain
\begin{eqnarray}
d^{LY}_{\phi_{(1,1)}}=1~, \ \ \ d^{LY}_{\phi_{(1,2)}}=\xi^{-3}+1+\xi^3~, \ \ \  T^{LY} = {\rm diag}(1,\xi^9) ~,
\end{eqnarray}
which is the Lee-Yang category. Note that both Lee-Yang and its complex conjugate have the same spectrum of quantum dimensions since
\begin{equation}
d^{LY}_{\phi_{(1,2)}}=\xi^{-3}+1+\xi^3=\xi^{-2}+1+\xi^2=d^{\overline{LY}}_{\phi_{(1,2)}}~.
\end{equation}

From this discussion, we see that the rational theory contains a sub-category that is Galois conjugate to the MTC for the IR chiral algebra in the flow discussed around \eqref{RGflowA} (the Lee-Yang or $(2,5)$ minimal model Virasoro algebra corresponding to $\chi[(A_1, A_2)]$). Therefore, the rational UV theory \lq\lq knows" about the IR MTC.

\begin{figure}
\begin{center}
\includegraphics[height=1.8in,width=4in,angle=0]{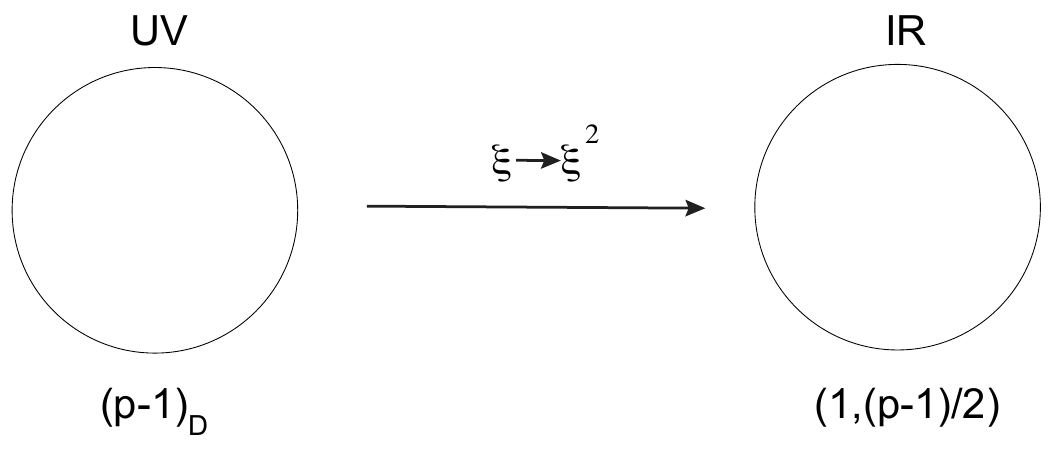}
\caption{The anyonic imprint on the Higgs branch. The expectation value for the Wilson lines corresponding to the Dynkin label $(p-1)$ fields in the rational 2D theory related to the UV $(A_1, D_p)$ SCFT are mapped, via Galois conjugation, to the expectation value for the Wilson line corresponding to the lowest scaling dimension primary in the $(2,p)$ minimal model related to the IR $(A_1, A_{p-3})$ SCFT on the Higgs branch.}
\label{fig1}
\end{center}
\end{figure}

More generally, one may ask if the MTC for the $\tilde D_{p+1}$ theory contains a closed fusion subcategory corresponding to the representations of the $(2,p)$ Virasoro algebra for $p\ge7$. It turns out that for general $p\in\mathbb{Z}_{\rm odd}$, the $\tilde D_{p+1}$ MTC does not have a non-trivial closed subcategory. However, we can partly generalize what happens for $p=5$ as follows. The vev of the Wilson line in the $\tilde D_{p+1}$ Chern-Simons theory that corresponds to the maximal spin representation (and therefore, via the correspondence discussed above, to the 4D Schur operators) is related, via Galois conjugation, to the vev of a Wilson line in the $(2,p)$ MTC corresponding to the 4D IR theory (see Fig. \ref{fig1}).\footnote{Note that we are not claiming the UV and IR MTCs are Galois conjugate. Indeed, the number of simple elements is different.} In other words
\begin{equation}\label{GCstatement}
\langle W_{D,p-1}^i\rangle={1\over2\sin\left({\pi\over2p}\right)}={S^D_{0,(p-1)_i}\over S^D_{0,0}}\ \ \rightarrow_{2\in\mathbb{Z}_p^{\times}} \ \ {S_{(1,1),(1,(p-1)/2)}\over S_{(1,1),(1,1)}}={(-1)^{p+1\over2}\over2\cos\left({\pi\over p}\right)}=\langle W_{(1,(p-1)/2)}\rangle~,
\end{equation}
where \lq\lq$\rightarrow_{2\in\mathbb{Z}_p^{\times}}$" denotes Galois conjugation by the element $2\in\mathbb{Z}_p^{\times}$ (since $p\in\mathbb{Z}_{\rm odd}$, this is always an element of the Galois group), and
\begin{equation}\label{2,pS-matrix1}
S_{(r,s),(\rho,\sigma)}={2\over\sqrt{p}}(-1)^{s\rho+r\sigma}\sin\left({\pi p\over2}r\rho\right)\sin\left({2\pi\over p}s\sigma\right)~.
\end{equation}
is the $(2,p)$ minimal model $S$-matrix \cite{DiFrancesco:1997nk}.\footnote{Note that, as in the $p=5$ example, the Galois conjugate of the $(p-1)_i$ twists generally do not agree with the twist for $(1,(p-1)/2)$ in the IR MTC. On the other hand, conjugating by $3\in\mathbb{Z}_p^{\times}$ (when $p$ is not a multiple of $3$) does yield an equality of the twists. However, for general $p$ not a multiple of three, we do not have a relation of quantum dimensions as in \eqref{GCstatement} if we choose $3\in\mathbb{Z}_p^{\times}$.} Note that the $\mathbb{Z}_p^{\times}$ Galois group we discuss here can be obtained from the appropriate surjective restriction of the $\mathbb{Z}_{2p}^{\times}$ Galois group (if $p-1=0\ {\rm mod}\ 4$) or $\mathbb{Z}_{4p}^{\times}$ Galois group (if $p-1=2\ {\rm mod}\ 4$) one finds by applying the discussion in \cite{Bantay:2001ni} to the full $\tilde D_{p+1}$ RCFT modular data (a similar statement holds for the Galois group that naturally arises when considering the underlying quantum group).

For the interested reader, we give the proof of \eqref{GCstatement} in Appendix \ref{app}. Here we mention a few  observations before discussing some generalizations in the next section:
\begin{itemize}
\item{The identification in \eqref{GCstatement} leads to some simple rules that one can easily verify for the theories in question. For example, if $\langle W^i_{D,p-1}\rangle\ne1$, then both the UV and the IR theory have non-abelian anyons. The reason is that such a quantum dimension cannot equal one when raised to any power and so the corresponding Wilson line / anyon cannot satisfy group-like fusion (this statement holds even though the IR theory is non-unitary, and the quantum dimension bound in \eqref{dimbound} is violated in the IR). Indeed, the $\tilde D_{p+1}$ and $(2,p)$ theories with $p>5$ have non-abelian anyons (in fact, any non-unitary MTC must have non-abelian anyons). When $\langle W^i_{D,p-1}\rangle=1$, the UV theory has an abelian anyon, and the IR (after removing the decoupled hypermultiplet) must also have an abelian anyon in its MTC or be trivial. As we have seen, the only such case in our theories is the $p=3$ case, where the UV has $\mathbb{Z}_3$ abelian anyons and the IR is trivial (after considering the theory related to the 4D IR in which we have removed the Goldstone multiplet). In the next section, we will comment on some generalizations of these observations to other theories.} 
\item{The quantum dimension on the LHS of \eqref{GCstatement} is related to the field in the $\tilde D_{p+1}$ theory whose character reproduces the Schur index of the UV $(A_1, D_p)$ SCFT. On the other hand, the quantum dimension on the RHS of \eqref{GCstatement} is related to the field whose character reproduces the Schur index of the IR $(A_1, A_{p-3})$ theory in the presence of an $\CN=(2,2)$-preserving surface defect \cite{Cordova:2017mhb}.} 
\item{It is interesting to note that in the MTCs that are related to our 4D $\CN=2$ SCFTs, all bosonically generated one-form symmetries (i.e., the corresponding generators have integer spin) have been gauged: the $\mathbb{Z}_2$ one-form symmetry in $SU(2)_{2(p-1)}$ has been gauged, and the corresponding bosons have condensed. In the $p=3$ case we have a left-over one-form symmetry, $\mathbb{Z}_3$, after the $\mathbb{Z}_2$ gauging (note that the anyons generating the $\mathbb{Z}_3$ symmetry have spin $1/3$). However, this symmetry has a 't Hooft anomaly---and hence cannot be gauged (e.g., see the recent discussion in \cite{Hsin:2018vcg}). 
}
\end{itemize}

\newsec{Connections with other theories}
It would be interesting to understand how general the observations in the previous section are in the space of 4D $\CN=2$ SCFTs. As a modest first step, let us revisit the $D_2[SU(3)]=(A_2^3[-1],F)$ SCFT\footnote{We use the language of section \ref{4DRGflow} in writing $(A_2^3[-1],F)$.} discussed in \cite{Buican:2017rya} and recounted briefly in the introduction. Recall from the introduction that the associated non-unitary chiral algebra in the sense of \cite{Beem:2013sza} is $\widehat{su(3)}_{-{3\over2}}$ \cite{Buican:2015ina} and that the associated unitary RCFT discussed in \cite{Buican:2017rya} is $\widehat{so(8)}_1$.\footnote{The 4D interpretation of this 2D unitary theory is as the ($\mathbb{Z}_2$ orbifold of the) free theory of eight non-unitary hypermulitplets with wrong spin-statistics.} The corresponding MTC is $Spin(8)_1$ (e.g., see the recent discussion in \cite{Seiberg:2016rsg}).

As in the examples mentioned in the previous sections, the $Spin(8)_1$ TQFT has no one-form symmetries generated by bosons. Indeed, all the non-trivial lines are fermionic. One can gauge a $\mathbb{Z}_2$ one-form symmetry generated by one of the fermions and obtain the $SO(8)_1$ spin-TQFT.\footnote{It might also be interesting to pursue ideas along the lines of \cite{Aasen:2017ubm}.}

More generally, as explained in footnote \ref{toricfootnote}, the results of \cite{Buican:2017rya} imply the following MTCs are associated with the $D_2[SU(2n+1)]=(A_{2n}^{2n+1}[1-2n],F)$ 4D $\CN=2$ theories
\begin{equation}\label{MTCD2}
D_2[SU(2n+1)]\ \rightarrow\ \begin{cases}
Spin(8)_1\ {\rm MTC}~, &\text{if}\ n(n+1)=2 \ {(\rm mod 4)} \\
D(\mathbb{Z}_2)\ {\rm (toric \ code)\  MTC}~, &\text{if}\ n(n+1)=0 \ {(\rm mod 4)}~.
\end{cases}
\end{equation}
The toric code MTC has two non-trivial bosons that can condense. However, this condensation leads to a trivial theory.\footnote{This statement follows, as in the related discussion around \eqref{su2S} for $SU(2)_{2(p-1)}$, from the modular $S$-matrix \cite{rowell2009classification} of the toric code MTC; see also \cite{neupert2016boson}.} Therefore, we see that all the MTCs that are related to the doubly infinite classes of 4D $\CN=2$ SCFTs discussed in the present paper do not allow for further non-trivial gauging of bosonic one-form symmetries. It would be interesting to understand if this is a general feature of MTCs related to 4D theories in the way we have described.

As in the case of the $(A_1, D_3)$ theory, the MTCs described in \eqref{MTCD2} are abelian: they have $\mathbb{Z}_2\times\mathbb{Z}_2$ fusion rules. Moreover, as for the $(A_1, D_3)$ theory, the Higgs branches of these theories at generic points are free: they consist of decoupled hypermultiplets (the would-be $A_{N-1}$ $W$-algebra minimal models in \eqref{chiralgs} do not exist, since $p<N$). Therefore, we see that, by again dropping decoupled hypermultiplets, UV and IR quantum dimensions can be related as in \eqref{GCstatement}\footnote{For $n=1$, it is natural to include $\langle W_{[1,0,0,0]}\rangle$ since this line corresponds to the $\widehat{so(8)}_1$ primary with (co-highest) conformal weight.}
\begin{equation}\label{D2match}
\langle W_{[0,0,\cdots,1]}\rangle=\langle W_{[0,0,\cdots,1,0]}\rangle=1={S_{0,[0,\cdots,1]}\over S_{0,0}}={S_{0,[0,\cdots,1,0]}\over S_{0,0}}\ \ \rightarrow_{1\in\mathbb{Z}_1^{\times}} \ \ {S_{00}\over S_{00}}=1=\langle W_{0}\rangle~,
\end{equation}
where the representations on the LHS correspond to the highest conformal weight primaries in the respective $\widehat{so(4n(n+1))}_1$ chiral RCFTs. As in the case of the $(A_1, D_3)$ theory, the quantum dimension on the RHS is for the trivial theory without an $\CN=(2,2)$-preserving surface defect included.

Before concluding, we should note an additional subtlety for the $D_2[SU(2n+1)]$ theories with $n>1$. In this case, we have non-generic flows to theories of the type $D_2[SU(2n'+1)]$ with $n'<n$ and decoupled hypermultiplets. As a result, we have interacting IR factors. However, as we have shown above, the related chiral RCFTs have only abelian anyons. Therefore, we again have a matching as in \eqref{D2match} if we also \lq\lq rationalize" the IR theory. The fact that the IR chiral RCFTs have only abelian anyons is consistent with our Galois action described above: the relevant Galois groups for $Spin(8)_1$ and $D(\mathbb{Z}_2)$ are trivial.

\newsec{Conclusions}
We conclude with some additional observations, comments, and open questions:
\begin{itemize}
\item{Another way to find a unitary interpretation of the $\chi[D_2[SU(3)]]=\widehat{su(3)}_{-{3\over2}}$ characters discussed around \eqref{so8charrel} and in this section is as follows. Consider the chiral $\widehat{su(3)}_3$ CFT. Three of the ten primaries of this theory (transforming under $su(3)$ representations $[0,0],[3,0],$ and $[0,3]$) are related to the abelian lines that generate the $\mathbb{Z}_3$ one-form symmetry of the $SU(3)_3$ MTC. Gauging this one-form symmetry projects out the lines in representations $[0,1], [1,2], [2,0], [0,2], [2,1], [1,0]$. The remaining $[1,1]$ representation is a fixed point under $\mathbb{Z}_3$ fusion, and so we add two more copies of it. This object then gives rise to the three unitary dimension $1/2$ chiral primaries in the associated chiral RCFT whose characters match the $\widehat{su(3)}_{-{3\over2}}$ vacuum character.

This approach is reminiscent of the $\mathbb{Z}_2$ gauging in the case of $\widehat{su(2)}_{2(p-1)}$ discussed at length in the present paper and also in \cite{Mukhi:1989bp}. As in the $\widehat{su(2)}$ case, it potentially gives us a canonical way to relate the unitary and non-unitary theories when we turn on (discrete) flavor fugacities. This example, combined with those in the rest of this paper, suggest a link between the physics of the $(A_{N-1}^N[p-N],F)$ theories, the admissible characters of their associated chiral algebras, $\widehat{su(N)}_{N{1-p\over p}}$, and the $\mathbb{Z}_{N(p-1)}$ gaugings of $\widehat{su(N)}_{N(p-1)}$. The relation is already somewhat more elaborate in the case of $N=2n+1\ge5$ and $p=2$, since the anyons related to the one-form symmetry correspond to 2D RCFT chiral primaries of conformal dimension larger than 1.}
\item{For general $p$ and $N$ one must also take into account the fact that some of the admissible characters of the logarithmic theories have negative coefficients.\footnote{This statement can be easily seen by considering the linear modular differential equations (LMDEs) satisfied by the Schur index (e.g., see \cite{Beem:2017ooy} for an introduction in the context of the 4D/2D correspondence of \cite{Beem:2013sza}). For other interesting recent work on LMDEs and their implications for 2D CFT, see \cite{Chandra:2018pjq,Bae:2018qym,Bae:2018qfh}.} Perhaps these can be related to rational theories after turning on some flavor fugacities (or, more generally, fugacities for generators corresponding to a unitary $W$-algebra). Clearly it would be interesting to understand this point better}
\item{In our examples, we \lq\lq rationalized" UV chiral CFTs constructed via \cite{Beem:2013sza} by associating rational theories with them. On the other hand, the IR was already rational, though non-unitary, at the $(A_1, A_{p-3})$ endpoints (since it was a $(2,p)$ minimal model). More generally, to allow for an anyonic imprint on the Higgs branch as in \eqref{GCstatement} and Fig. \ref{fig1}, we have to \lq\lq rationalize" the IR theory as well. Indeed, we saw an example of this phenomenon in the $D_2[SU(2n+1)]\to D_2[SU(2n'+1)]$ flows. It would be interesting to understand this process more generally.}
\item{Our most non-trivial correspondence (i.e., the one with a non-trivial Galois action) was between UV chiral RCFTs and IR chiral algebras that are $C_2$-cofinite. In physics language, this means that we are studying IR theories on the Higgs branch that have no Higgs branches themselves \cite{Beem:2017ooy} (e.g., the $(A_1, A_{p-3})$ theories with $p\in\mathbb{Z}_{\rm odd}$ do not have Higgs branches). The authors of \cite{Beem:2017ooy} and their collaborators have embarked on a program to classify 4D $\CN=2$ SCFTs using these $C_2$-cofinite theories as basic building blocks. It would be interesting if our work sheds light on this program.}
\item{We did not pursue qDS reduction on the RCFT side. Clearly this is interesting to do. Perhaps the recent notion of Galois conjugation at the level of RCFT characters \cite{Harvey:2018rdc} will prove useful to make contact between the UV and IR. The LMDE-based discussion in \cite{Chandra:2018pjq} may also play a role.}
\end{itemize}
We hope to return to some of these questions soon.

\ack{We are particularly grateful to T.~Nishinaka for many interesting discussion on these and related topics. We also thank R.~Radhakrishnan and S.~Wood for interesting comments and discussions. M. B.'s research is partially supported by the Royal Society under the grant \lq\lq New Constraints and Phenomena in Quantum Field Theory." Z. L. is supported by a Queen Mary University of London PhD studentship.}

\newpage
\begin{appendices}
\section{Proof of \eqref{GCstatement}}\label{app}
In this appendix, we will prove \eqref{GCstatement}. For ease of reference, we reproduce it below
\begin{equation}\label{GCstatementA}
\langle W_{D,p-1}^i\rangle={1\over2\sin\left({\pi\over2p}\right)}={S^D_{0,(p-1)_i}\over S^D_{0,0}}\ \ \rightarrow_{2\in\mathbb{Z}_p^{\times}} \ \ {S_{(1,1),(1,(p-1)/2)}\over S_{(1,1),(1,1)}}={(-1)^{p+1\over2}\over2\cos\left({\pi\over p}\right)}=\langle W_{(1,(p-1)/2)}\rangle~.
\end{equation}
To obtain these elements we will use the S-transformation properties of the $\widehat{su(2)}_{2(p-1)}$ primaries given by the S-matrix in \eqref{su2S} which we reproduce below
\begin{equation}\label{uncondensed_S}
  S_{l_{1},l_{2}} = \frac{1}{\sqrt{p}} \sin \left[ \frac{(l_{1} + 1)(l_{2} + 1) \pi}{2p} \right] \ . 
\end{equation}
As discussed in \eqref{Dreps}, primaries of the condensed $\tilde D_{p+1}$ theory take the following form in terms of primaries of $\widehat{su(2)}_{2(p-1)}$
\begin{equation}\label{DrepsA}
\Phi_{D,\ell}=\Phi_{\ell}\oplus\Phi_{2(p-1)-\ell}~, \ \ \ \ell\in\{0,2,4,\cdots,p-3\}~, \ \ \ \Phi_{D,p-1}^i=\Phi_{p-1}^i~,
\end{equation}
where $i=1,2$. To calculate elements of the first row of the $\tilde D_{p+1}$ S-matrix we need to write the S-transformation of the condensed vacuum in terms of condensed fields using \eqref{uncondensed_S} 
\begin{eqnarray}
(S^D\chi_{D})_0&=&\sum_{\ell=0, \ell\in\mathbb{Z}_{\rm even}}^{p-3}S^D_{0,\ell}\chi_{D,\ell}+\sum_{i=1}^2S^D_{0,(p-1)_i}\chi_{D,(p-1)_i}=\sum_{\ell=0}^{2(p-1)}(S_{0,\ell}+S_{2(p-1),\ell})\chi_{\ell}^{\widehat{su(2)}_{2(p-1)}}\cr&=&\sum_{\ell=0}^{2(p-1)} \frac{1}{\sqrt{p}} \sin\left( \frac{\ell+1}{2p} \pi\right) \chi_{\ell}^{\widehat{su(2)}_{2(p-1)}} + \sum_{\ell=0}^{2(p-1)} \frac{1}{\sqrt{p}} \sin \left(\frac{(2p-1)(\ell+1)}{2p} \pi \right)\chi_{\ell}^{\widehat{su(2)}_{2(p-1)}}\cr&=&\sum_{\ell=0}^{p-1}{2\over\sqrt{p}}\sin\left({2\ell+1\over2p}\pi\right)\chi_{2\ell}^{\widehat{su(2)}_{2(p-1)}}~.
\end{eqnarray}
In going to the last equality, we have used the relation $\sin\left(\frac{(2p-1)(\ell+1)}{2p} \pi\right) = (-1)^{\ell} \sin\left(\frac{\ell+1}{2p} \pi\right)$.

Now, we can solve for the first $(p-1)/2$ entries of the first row of the $S^D$ matrix
\begin{equation}
S^D_{0,\ell}={2\over\sqrt{p}}\sin\left({2\ell+1\over2p}\pi\right)~.
\end{equation}
The last two entries of the first row are also constrained to obey
\begin{equation}
S^D_{0,(p-1)_1}+S^D_{0,(p-1)_2}={2\over\sqrt{p}}~, \ \ \ S^D_{0,(p-1)_1}\in\mathbb{R}~,
\end{equation}
where the reality of these entries is required by the reality of the quantum dimensions. Unitarity of the $S$-matrix requires the first row to have unit norm and so
\begin{equation}
S^D_{0,(p-1)_1}=S^D_{0,(p-1)_2}={1\over\sqrt{p}}~.
\end{equation}
In particular, we see that the quantum dimension in the UV theory is indeed
\begin{equation}
{S^D_{0,(p-1)_i}\over S^D_{0,0}}={1\over2\sin\left({\pi\over2p}\right)}~,
\end{equation}
as claimed in \eqref{GCstatementA}.

Now let us study the quantum dimension of $\phi_{(1,(p-1)/2)}$. This quantity is easily computed from the $(2,p)$ S-matrix
\begin{equation}\label{2,pS-matrix}
S_{(r,s),(\rho,\sigma)}={2\over\sqrt{p}}(-1)^{s\rho+r\sigma}\sin\left({\pi p\over2}r\rho\right)\sin\left({2\pi\over p}s\sigma\right)~.
\end{equation}
Indeed, we find
\begin{equation}
{S_{(1,1),(1,(p-1)/2)}\over S_{(1,1),(1,1)}}={(-1)^{p+1\over2}\over2\cos\left({\pi\over p}\right)}~,
\end{equation}
as claimed in \eqref{GCstatementA}.

Now we would like to discuss the Galois action that relates the two quantum dimensions. First, we claim that the Galois group acting on the quantum dimensions (and also the $T$ matrices) can be taken to be $G=\mathbb{Z}_p^{\times}$ (see the main text for a discussion of the reduction to $G$ from the larger groups one finds using the methods of \cite{Bantay:2001ni} and also from the underlying quantum groups). For the $T$ matrices (defined with the normalization in \eqref{Tmatdef}), this statement follows from \eqref{su2Udims} since $\ell\in\mathbb{Z}_{\rm even}$ and so the $\theta_{\ell}$ are $p^{\rm th}$ roots of unity (a similar statement holds on the $(2,p)$ minimal model side). At the level of the quantum dimensions, it is sufficient to show that $\sin\left({2\ell+1\over2p}\pi\right)$ can be written in the field $\mathbb{Q}(\xi)$, where $\xi=e^{2\pi i\over p}$.

To see this statement is correct, note that since $p$ is odd, we have either $p+2\ell+1=4n_{\ell}$ or $p+2\ell+1=4n_{\ell}+2$ for $n_{\ell}\in\mathbb{Z}$. In either case, we have
\begin{equation}
\sin\left({2\ell+1\over2p}\pi\right)={1\over2} \left(e^{\frac{i \pi}{2} \left(\frac{2\ell+1}{p} - 1\right)} + e^{-\frac{i \pi}{2} \left(\frac{2\ell+1}{p} - 1\right)}\right) \ .
\end{equation}
Let us now suppose $p+2\ell+1=4n_{\ell}$. We then have
\begin{eqnarray}
\sin\left({2\ell+1\over2p}\pi\right)&=&-{1\over2}\left({e^{\frac{\pi i}{2}\left(\frac{2\ell+1}{p} + 1\right)} + e^{-\frac{\pi i}{2}\left(\frac{2\ell+1}{p} + 1\right)}}\right)=-{1\over2}\left({e^{\frac{2\pi in_{\ell}}{p}} + e^{-\frac{2\pi in_{\ell}}{p}}}\right)\cr&=&{(-1)^{p-1\over2}\over2}\left(\xi^{n_{\ell}}+\xi^{-n_{\ell}}\right)\in\mathbb{Q}(\xi)~,
\end{eqnarray}
as desired. Similarly, for $p+2\ell+1=4n_{\ell}+2$, we have
\begin{equation}
\sin\left({2\ell+1\over2p}\pi\right)={1\over2}\left({e^{\frac{2\pi i(\ell-n_{\ell})}{p}} + e^{-\frac{2\pi i(\ell-n_{\ell})}{p}}}\right)={(-1)^{p-1\over2}\over2}\left(\xi^{n_{\ell}-\ell}+\xi^{\ell-n_{\ell}}\right)\in\mathbb{Q}(\xi)~,
\end{equation}
which completes our proof of the claim that $G=\mathbb{Z}_p^{\times}$.

Let us now apply the Galois action $2\in G$ to the unitary quantum dimension. We have from the previous two equations that
\begin{equation}
  \frac{1}{2 \sin \frac{\pi}{2p}} = \frac{(-1)^{\frac{p-1}{2}}}{\xi^{n} + \xi^{-n}} \ , \ \ \ n =  \left  \lfloor \frac{p+1}{4} \right \rfloor~.
\end{equation}
Now, applying the Galois action yields
\begin{equation}
\frac{1}{2\sin \frac{\pi}{2p}} = \frac{(-1)^{\frac{p-1}{2}}}{\xi^{n} + \xi^{-n}} \longrightarrow  \frac{(-1)^{\frac{p-1}{2}}}{\xi^{2n} + \xi^{-2n}} = \frac{(-1)^{\frac{p-1}{2}}}{2\cos \left(\frac{4\pi n}{p}\right)} = \frac{(-1)^{\frac{p+1}{2}}}{2\cos \left(\frac{\pi }{p}\right)} \ ,
\end{equation}
where in the last equality we used the relation $\cos \left(\frac{4 \pi n}{p}\right) = - \cos \left(\pi \frac{4n - p}{p}\right) = - \cos \frac{\pi}{p}$ for $p = 4n \pm 1$. This completes the proof of our assertion in \eqref{GCstatementA} / \eqref{GCstatement}.

\end{appendices}

\newpage
\bibliography{chetdocbib}

\end{document}